# White Paper - Objectionable Online Content: What is harmful, to whom, and why

Thamar Solorio[1], Mahsa Shafaei[1], Christos Smailis[1], Brad J. Bushman[2], Douglas A. Gentile[3], Erica Scharrer[4], Laura Stockdale[5], Ioannis Kakadiaris[1]

[1]Department of Computer Science, University of Houston
[2]School of Communication, The Ohio State University
[3]Department of Psychology, Iowa State University
[4]Department of Communication, University of Massachusetts Amherst
[5]Department of Family Life, Brigham Young University

**Executive Summary**

This White Paper summarizes the authors' discussion regarding objectionable content for the University of Houston (UH) Research Team to outline a strategy for building an extensive repository of online videos to support research into automated multimodal approaches to detect objectionable content. The workshop focused on defining what harmful content is, to whom it is harmful, and why it is harmful.

## A. BACKGROUND

Consuming mass media online, whether it is television, movies, videos, or video games, made widely available by the multiple streaming apps (e.g., YouTube, Netflix, Apple TV, Hulu, Amazon Prime, TikTok, Facebook Live), has become a prime form of entertainment for people of all ages. These mass media platforms usually offer harmless content that, in many cases, can be educational, such as the TV program Sesame Street, which has been shown to improve social and reading skills in young children [1]. or the PBS show Daniel Tiger's neighborhood [2], [3]. However, there is also strong evidence showing the many adverse effects of media on young viewers, that range from instigating aggressive behavior [4], [5], perpetuating (sex, race, age)-ism mentality, irresponsible sexual or alcohol consumption, body image-related concerns, concerns about outcomes about sex stated separately from "irresponsible sexual consumption," concerns about health effects stemming from the marketing of low nutrient, high salt, high sugar, high-fat content foods, commercialization, and other violence-related outcomes such as desensitization, the normalization of violence and inducing anxiety and fear.

Rating mechanisms of online content are aimed at providing parents/guardians with information regarding the type of content included in some of these videos. However, as it was made evident by incidents such as Elsagate, where adult content was presented as targeting children with the use of characters from the well-known Disney Movie Frozen, filtering provided by some streaming platforms is lacking accuracy reliability. Alternative mechanisms that rely on human judges, such as the ones used by the Motion Picture Association of America (MPAA), are not scalable to cover the amount of content distributed online platforms. *There is an urgent need to develop automated approaches to provide content labels that can support parents', guardians', and educators' decision process on whether they want to allow their kids to access specific content or not.*

The nature of online material, where objectionable content can be expressed on any single or combination of different modalities (audio, images, text), motivates the need to develop a repository with annotations covering each modality. Additionally, because objectionable content is an umbrella term to cover several types of content that can be considered harmful, they can also have different degrees of severity. There is a need to create a taxonomy to detect objectionable content and that will support the annotation process. In what follows, we discuss different aspects relevant to developing a repository that can effectively support R&D in this space.

**B. TOWARDS A TAXONOMY OF DETECT OBJECTIONABLE CONTENT**

To develop a taxonomy, we must understand the different types of content that the fields of psychology and communication have identified as having a negative impact on viewers, especially children. The authors identify the following significant types of content that need to be considered:

a. *Violence:* in all its different forms. For example, associating masculinity with violence and guns. Live stream videos with raw and graphic content [1] [2] [3] [4] [5]. Consider the live stream events of past mass shootings, acts of terrorism, and brutal demonstrations of power, all of these videos are distributed live without any content warning. In addition, we have content that is glamorizing/glorifying violence (done by likable characters, presented as justified, rewarded in the narrative, or at least not punished) or minimizing violence (not showing realistic pain or harm, glossing over consequences, combined with humor).

b. *Sexual messages:* the main concern with pornography is not so much about early sexual exploration in teens but about the gender messages available in these videos and the amount of aggression in mainstream adult films [6] [7]. Regarding other sexual messages in the media, additional concerns related to lack of depiction of health risks (such as STDs/STIs) or contraception and with the ability of exposure to media with sexual messages to shape norms associated with sex (such as perceptions of whether peers are sexually active) which, in turn, can shape one's sexual behavior.

c. *The pairing of pleasure with violence.* What is really dangerous is when the content is associating something pleasant (e.g., sex, humor) with violence. Based on classical conditioning, when you pair sex and violence, violence becomes sexually arousing. For example, when you pair violence with humor, violence seems funny, which it is not, and the outcome can be perilous [5].

d. *Misguided meta-messages in media:* The message often conveyed in the media is that if something feels bad, it means it is wrong. For example, if people experience some very normal feelings (being anxious), they are supposed to do something or buy something (e.g., a drug). It fosters a culture of detaching from feelings and replacing them with material objects or looking for a solution in medications, leaving out the possibility for introspection. Other outcome variables studied under commercialism include the desire for less healthy food or other products advertised to kids, purchase requests/nagging, materialism, feelings of deprivation/dissatisfaction. This is an especially important area developmentally because studies show that it takes some time to develop the ability to identify marketing messages in digital spaces and to know they are attempts as persuasion. There is a missed opportunity for self-growth due to the constant distraction of consumerism culture [8] [9] [10] [11].

e. Verbal aggression, callousness, social aggression*:* Another harmful impact of the media is propagating lower levels of aggression. Nowadays, in almost every sitcom series, the story is not about the situation; instead, it is about relationships. For example, in TV series such as "Friends" or "Big Bang Theory," how people treat each other is the source of humor. Sarcasm, which is a form of aggression, takes a central role. It consists of putting down others and making fun of each other. It normalizes the idea that if you want to be friendly, you should be mean to everyone around you. That is the beginning of what later becomes toxic masculinity. This behavior is particularly concerning when young viewers are receiving this message [12] [2].

Based on the points highlighted above, we propose a taxonomy of objectionable content, as shown in Figure 1. The idea is that parent nodes represent a more general category of objectionable content, whereas child nodes represent more specific types of objectionable content. The taxonomy only captures the most relevant types of objectionable content; it does not address the severity of the content in its current form. We will discuss that later.

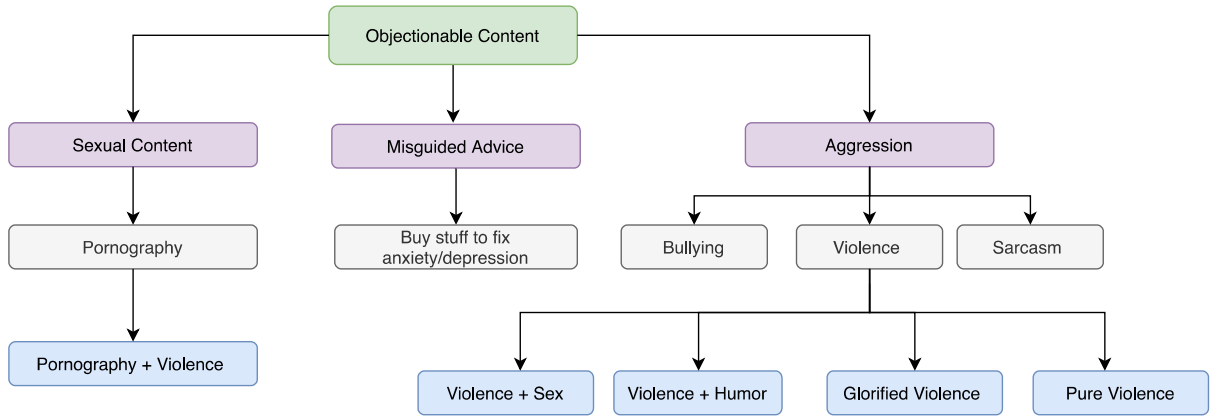

Figure 1. Taxonomy of objectionable content. Our latest taxonomy can be found at [arXiv].

### C. Are There Different Levels of Severity in Objectionable Content?

There are different aspects to consider here. There is consensus with the respect that anything that can be criminal is deemed severe, for example, abuse or assault. But another relevant dimension is how the content is presented. More specifically, is the aggression or violence presented in a glamourized manner? In other words, how much violence exists in the content, and how is violence depicted? Typically, war movies include a lot of graphic violence. Still, most of them also show violence as something serious with dire consequences. In this respect, it is not high-risk content compared to a superhero movie where heroes are shown as having no negative consequences from the violence they impart, as long as it is targeted to evil enemies (i.e., violence is portrayed as justified). In that sense, the content may more easily encourage or elicit change in how humans behave in society to the point of defining new rules. Although severity is important, we believe that this should be addressed in a second iteration of the annotation. We recommend addressing the detection of most of the types of objectionable content included in the Taxonomy from Figure 1. Still, the severity level is best addressed after we can detect the presence/absence of objectionable content. We view severity as a continuous dimension rather than as a dichotomy (i.e., serious vs. not serious).

### D. Harms Imparted by Different Types of Objectionable Content

Research shows that objectionable content desensitizes audiences, influences behavior, leads to opportunities lost and affects how we view other people. In the following, we explore each item in detail.

1. *Desensitization.* The issue is not that every person that watches violent content, for example, will become violent; those cases happen but are rare. The main problem is that almost everyone who consumes violent media becomes desensitized or numb to others' pain and suffering [13] [5]. We have many things in our culture that desensitize people; media is one essential desensitizer that, as a society, we can have control over.
2. *Influence in behavior.* As mentioned above, watching violent content does not necessarily make people violent, but it can make people rude, affecting how they interact with others. It influences how viewers perceive the world; they may consider the world as a hostile place, misinterpreting others' intentions and actions and promoting defensive behavior [4] [9] [14]. Media also influence the viewer's perception of women, who are often depicted in sexualized and stereotypical ways. This encourages a perception of women as sex objects rather than human beings [7] [15]. Additionally, the message of how people treat each other is relevant. Children can improve their empathy skills and become kinder if they see people who care about each other. If they see people acting aggressively or being unkind, they also learn how to be less empathetic and less kind [16].
3. *Lost opportunities.* The amount of time in front of a screen is also an important aspect to be considered. We need to weigh in the opportunity costs of what viewers are not doing when they decide to spend time watching online media content. Not engaging in physical activities, for example, can be associated with physical health problems connected to childhood obesity in heavy media users [17] [18].
4. *Perception of self and others.* Another important aspect of media is the message we get about ourselves and others. In the news, for example, Islam is often linked to terrorism. Because Christians usually don't know Muslims in their own lives, they rely on news construction or narrow stereotypes portrayed in mass media platforms. They share the religious identity that they learned from the media with millions of people worldwide with a limited form of violence [19]. Another example is what preconceived notions/judgments come to mind depending on how a person looks like? The incorrect idea that people can be judged based on their appearance can result in alienation and a fragmented society [9] [14] [12] [15].

**E. On the Specific Form of Media/Platform**

Our conclusion is that there is no considerable difference in terms of specific platforms when the users are only viewing content and not engaging with it, as in the form of video games. Perhaps a more relevant question is then which platforms are more popular with our main target population of interest.

**F. Discussion**

The development of automated technology to detect objectionable content online is an urgent need that must be addressed to protect vulnerable populations. Our discussion aired the fact that objectionable content presents itself in explicit and implicit forms. Explicit forms are in the form of violent graphic content, sexually explicit content, and language, to name a few. Implicit content includes meta-messages that are embedded in the media, for example, sitcom series that associate sarcasm with humor, where the content influences viewers' behavior and attitudes. We will consider annotating content in two stages, explicit content first and then implicit content.

Another relevant aspect to consider is the identification of pro-social content. So far, we have focused our attention on objectionable content, but of equal relevance to parents/guardians is the ability to detect content that can positively affect viewers. This could include educational content, content that inspires viewers to be compassionate and kind to one another, as well as content that deters negative stereotyping.

**G. Conclusions**

In this paper, we have presented a summary of the findings of the first workshop on understanding objectionable content. As a result of the discussions, we have proposed an initial version of a taxonomy of content that can be considered as potentially having negative effects on viewers. This taxonomy can be used to guide the design of annotation standards and guidelines to support the construction of a large repository of objectionable content online.

**Acknowledgment:** This work was supported in part by the National Science Foundation under Grant No. IIS-2036368. Any opinions, findings, and conclusions, or recommendations expressed in this material are those of the authors and do not necessarily reflect the views of the National Science Foundation.